\documentstyle[epsfig]{mn}

\newcommand{\apj}{ApJ}

\newcommand{\bc}{\begin{center}}
\newcommand{\ec}{\end{center}}
\title[Galaxies and subhaloes in $\Lambda$CDM galaxy clusters]{Galaxies and
  subhaloes in $\Lambda$CDM galaxy clusters}

\author[L.~Gao, et al.]{L.~Gao$^1$\thanks{Email: gaoliang@mpa-garching.mpg.de},
  G.~De~Lucia $^1$, S.~D.~M.~White $^1$, A.~Jenkins $^2$
  \\
  $^1$Max--Planck--Institut f\"ur Astrophysik, D-85748 Garching, Germany \\
  $^2$Institute for Computational Cosmology, Department of Physics,University of Durham,South Road, Durham DH1 3LE, U.K.}

\begin{document}
\label{firstpage} \maketitle
\begin{abstract}
  We combine $10$ high resolution resimulations of cluster--sized dark haloes
  with semi--analytic galaxy formation modelling in order to compare the
  number density and velocity dispersion profiles of cluster galaxies with
  those of dark matter substructures (subhaloes). While the radial
  distribution of galaxies follows closely that of the dark matter, the
  distribution of dark matter subhaloes is much less centrally concentrated.
  The velocity dispersion profiles of galaxies are also very similar to those
  of the dark matter, while those for subhaloes are biased high, particularly
  in the inner regions of clusters. We explain how these differences, already
  clearly visible in earlier work, are a consequence of the formation of
  galaxies at the centres of dark matter haloes.  Galaxies and
  subhaloes represent \emph{different} populations and are not directly
  comparable.  Evolution produces a complex and strongly position--dependent
  relation between galaxies and the subhaloes in which they reside.  This
  relation can be properly modelled only by appropriate physical
  representation of the galaxy formation process.
\end{abstract}


\section{Introduction}
\label{sec:intro}

A variety of observational indicators have recently converged to establish the
$\Lambda$CDM cosmogony as the {\it de facto} standard model for the formation
of structure in our universe (e.g. Spergel et al 2003). For the general class
of such hierarchical models, Navarro, Frenk \& White (1996, 1997) showed that
the radial density profiles of nonlinear structures such as galaxy or cluster
dark haloes are well represented by a simple fitting formula of ``universal
shape''. As new galaxy surveys have amassed homogeneous data for large samples
of clusters, the mean radial profiles of both number density and velocity
dispersion have been found to conform quite closely to these NFW predictions
for the dark matter (Carlberg et al. 1997; Biviano \& Girardi 2003). Models
which follow galaxy formation and cluster assembly explicitly do reproduce such
parallel galaxy and dark matter profiles, even though the relation between the
luminosity and dark matter mass of individual galaxies shows a lot of scatter
and is predicted to depend strongly on clustercentric distance (Diaferio et al.
2001; Springel et al 2001).

The high resolution achieved by numerical simulations in recent years has
allowed detailed study of the properties of dark matter substructure
(subhaloes) within dark haloes (Tormen 1997; Ghigna et al. 1998, 2000;
Klypin et al. 1999a, 1999b; Stoehr et al. 2002, 2003; De Lucia et
al. 2004a; Diemand et al. 2004; Gill et al. 2004a, 2004b; Gao et
al. 2004).  These studies agree quite well on the structure, abundance
and radial distribution predicted for subhaloes, once  differences in
numerical resolution are accounted for. On the other hand, a variety
of contradictory conclusions have been drawn from comparison with the
observed properties of luminous objects in galaxy haloes and galaxy clusters
(compare the discussions in Moore et al. 1999; Klypin et al. 1999a;
Stoehr et al 2002, 2003; D'Onghia \& Lake 2003; Desai et al. 2003;
Diemand, Moore \& Stadel 2004;  Kravtsov et al. 2004; Willman et
al. 2004). We argue below that these disagreements can in most cases
be traced to insufficiently careful modelling of the relation between the
properties of subhaloes and those of the galaxies they contain.

In particular, a number of recent studies have noted that the radial
distribution of subhaloes within dark haloes is very shallow compared both to
that of the dark matter and to that of observed galaxies in real clusters
(Ghigna et al. 2000; De Lucia et al. 2004a; Diemand et al. 2004; Gill
et al 2004a; Gao et al 2004). Some of these authors concluded that
this difference may indicate a fundamental problem for the
$\Lambda$CDM model, failing to notice that the earlier simulations of
Springel et al. (2001) had followed substructure with comparable
numerical resolution and showed that modelling baryonic processes can
produce a galaxy profile in good agreement with observation. This
suggests there are serious inadequacies in a simple model where the
luminosity (or kinematics) of a galaxy is simply related to the mass
(or potential well depth) of the corresponding subhalo in a
dark--matter--only simulation. With the assumptions of Springel et
al. (2001) the relation between these properties shows very large
scatter and depends systematically on radius within a cluster
halo. This is because the stellar mass of galaxy is determined
primarily by its halo mass at the time the stellar component was
assembled rather than by its halo mass at the present day.

Semi--analytic models of the kind used by Springel et al. (2001) are an ideal
tool to explore the relation between dark matter subhaloes and the galaxies
they host. In this paper we use the improved semi--analytic model developed by
De Lucia, Kauffmann \& White (2004b) which is able to reproduce the observed
luminosity functions, metallicities and colour--magnitude relations of cluster
galaxies, as well as the metal content of the intracluster medium. We apply
this model to a set of ten high resolution dark--matter--only resimulations of
cluster formation in a $\Lambda$CDM universe, eight of which are also analysed
in companion papers on the systematic properties of subhalo populations in
$\Lambda$CDM dark haloes (Gao et al. 2004) and on the assembly of the central
cusps of $\Lambda$CDM clusters (Gao et al. 2003)

This Letter is structured as follows. In Sec.~\ref{sec:simsam}, we briefly
describe the simulations and the semi--analytic model used for this study.  In
Sec.~\ref{sec:analysis}, we study the spatial distributions and the velocity
dispersion profiles of galaxies and dark matter substructures and we explain
the differences between them. A discussion and a summary of our results are
presented in Sec.~\ref{sec:summary}.
\section{The Simulations and the Semi--analytic Model}
\label{sec:simsam}
We use a set of ten $N$--body resimulations of the formation of a massive
galaxy cluster in a $\Lambda$CDM Universe. The clusters range in mass $M_{200}$
from $4.5\times 10^{14}\,h^{-1}{\rm M_\odot}$ to $8.5 \times
10^{14}\,h^{-1}{\rm M_\odot}$, and were initially identified in a cosmological
simulation of a region $0.479\, h^{-1}$Gpc on a side (Yoshida, Sheth
\& Diaferio 2001) and run with numerical parameters suggested by Power
et al. (2003). Many of them have been studied previously Gao et
al. (2003, 2004) and Navarro et al. (2003). These resimulations were
carried out using the publicly available parallel $N$--body code
{\small GADGET} (Springel, Yoshida \& White 2001) with a particle mass of $5.12
\times 10^8\,h^{-1}{\rm M_\odot}$ and a force softening of
$\epsilon=5\,h^{-1}$kpc.  The cosmological parameters assumed were:
$\Omega_0=0.3$, $\Omega_{\Lambda}=0.7$, $h=0.7$ (we adopt the standard
convention $H_0=100\,h\, {\rm km \, s^{-1}Mpc^{-1}}$) and normalisation
$\sigma_8=0.9$. Dark matter subhaloes are identified using the algorithm
{\small SUBFIND} (Springel et al. 2001). All subhaloes containing at least $10$
particles are tracked. The numerical data for each simulation are stored at 50
times logarithmically space between $1+z=40$ and $1+z=1$ and tree structures
are built to follow the formation and merger history of each halo and its
subhaloes.

We follow the baryonic evolution using the semi--analytic model by De Lucia et
al. (2004b). As in Springel et al. (2001), the model explicitly follows the
evolution of the dark matter halo within which a galaxy forms, even after this
halo is accreted by a larger object and becomes one of its subhaloes.  The
model also follows the chemical and photometric evolution of cluster galaxies
in a self--consistent way, together with the chemical enrichment of the
intracluster medium.  De Lucia et al.  (2004b) have shown that their model
agrees with a large body of observational results for galaxies in the local
Universe, both in clusters and in the field.  We refer to the original paper
for a more detailed description. In this study, we use their `feedback' model
which they find to be the only one able to reproduce the observed decline in
baryon fraction from rich clusters to galaxy groups.

\section{Number density and velocity dispersion profiles for galaxies
and subhaloes}
\label{sec:analysis}

\begin{figure*}
\resizebox{8.5cm}{!}{\includegraphics{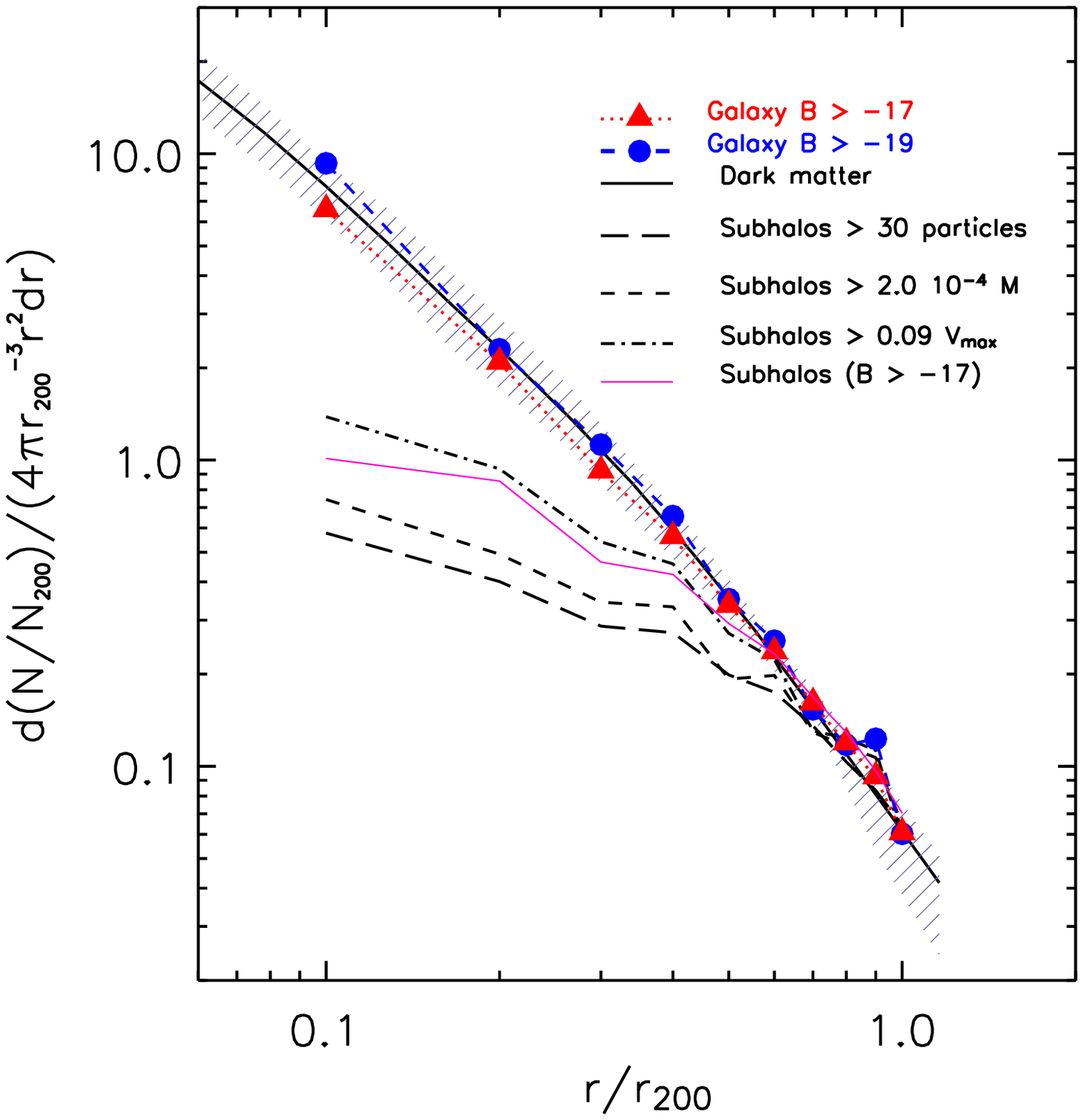}}\hspace{0.0cm}%
\resizebox{8.5cm}{!}{\includegraphics{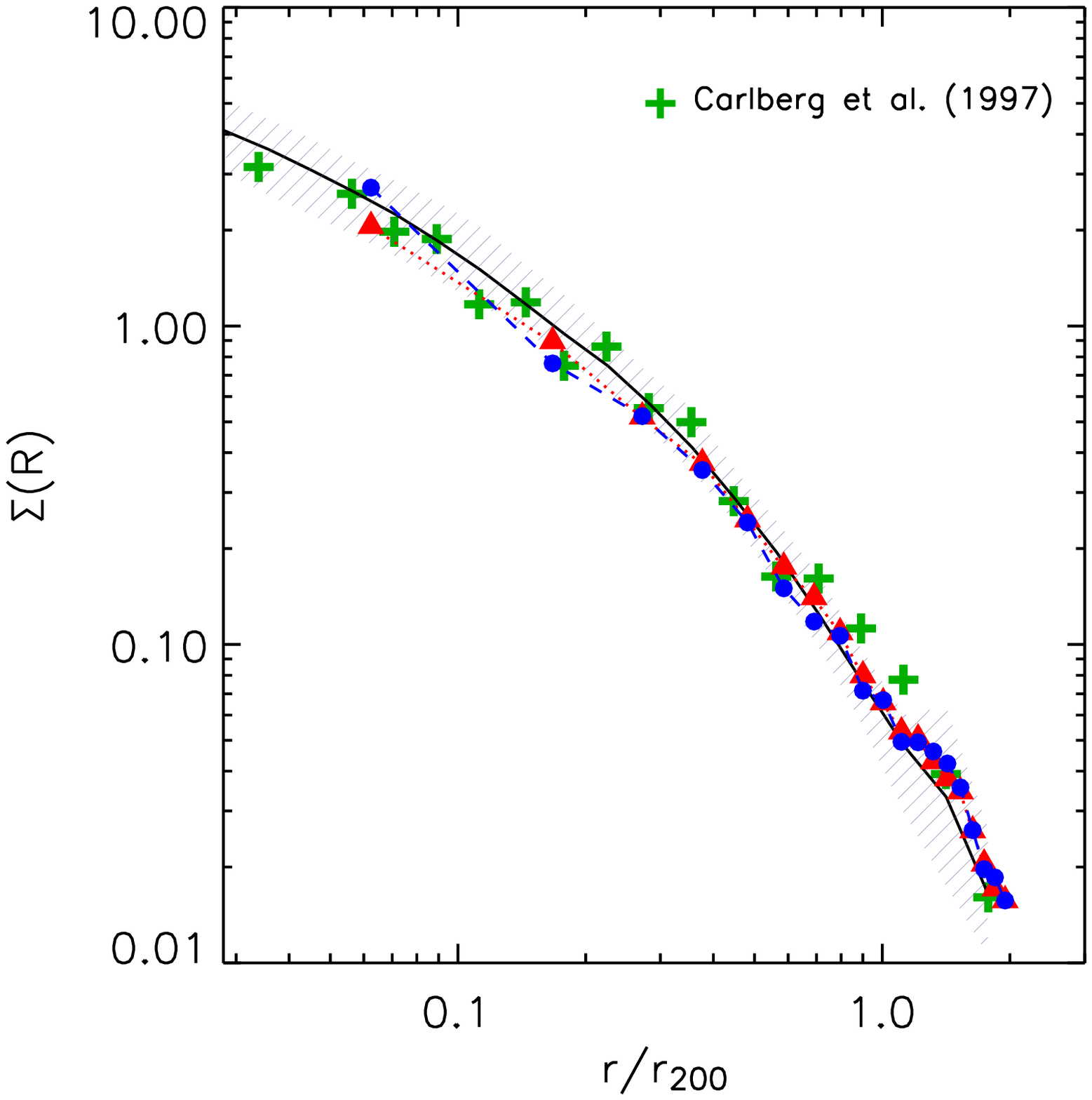}}\hspace{0.0cm}%
\vspace{-0.5cm}\
\resizebox{8.5cm} {!}{\includegraphics{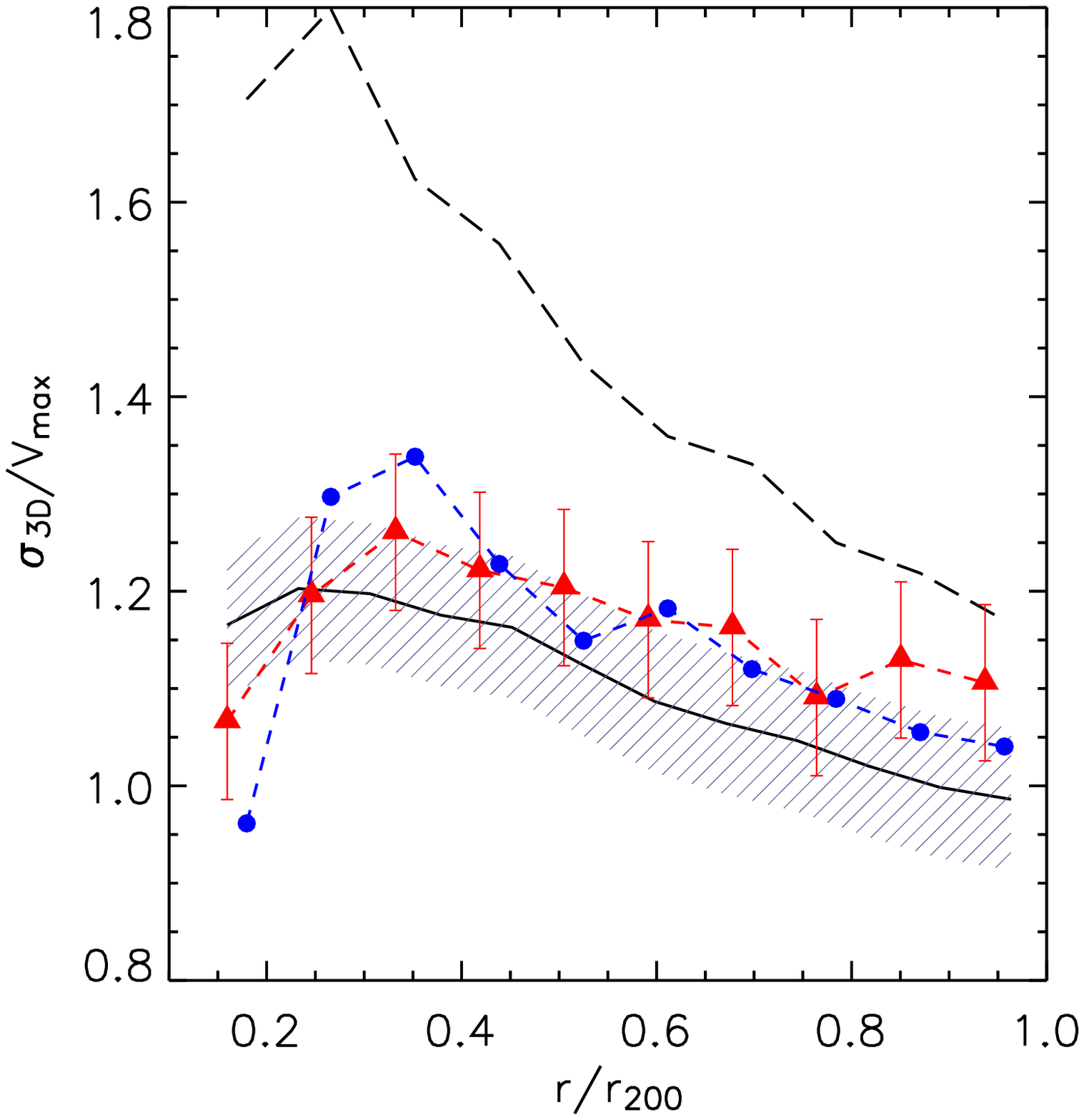}}\hspace{0.0cm}%
\resizebox{8.5cm}{!}{\includegraphics{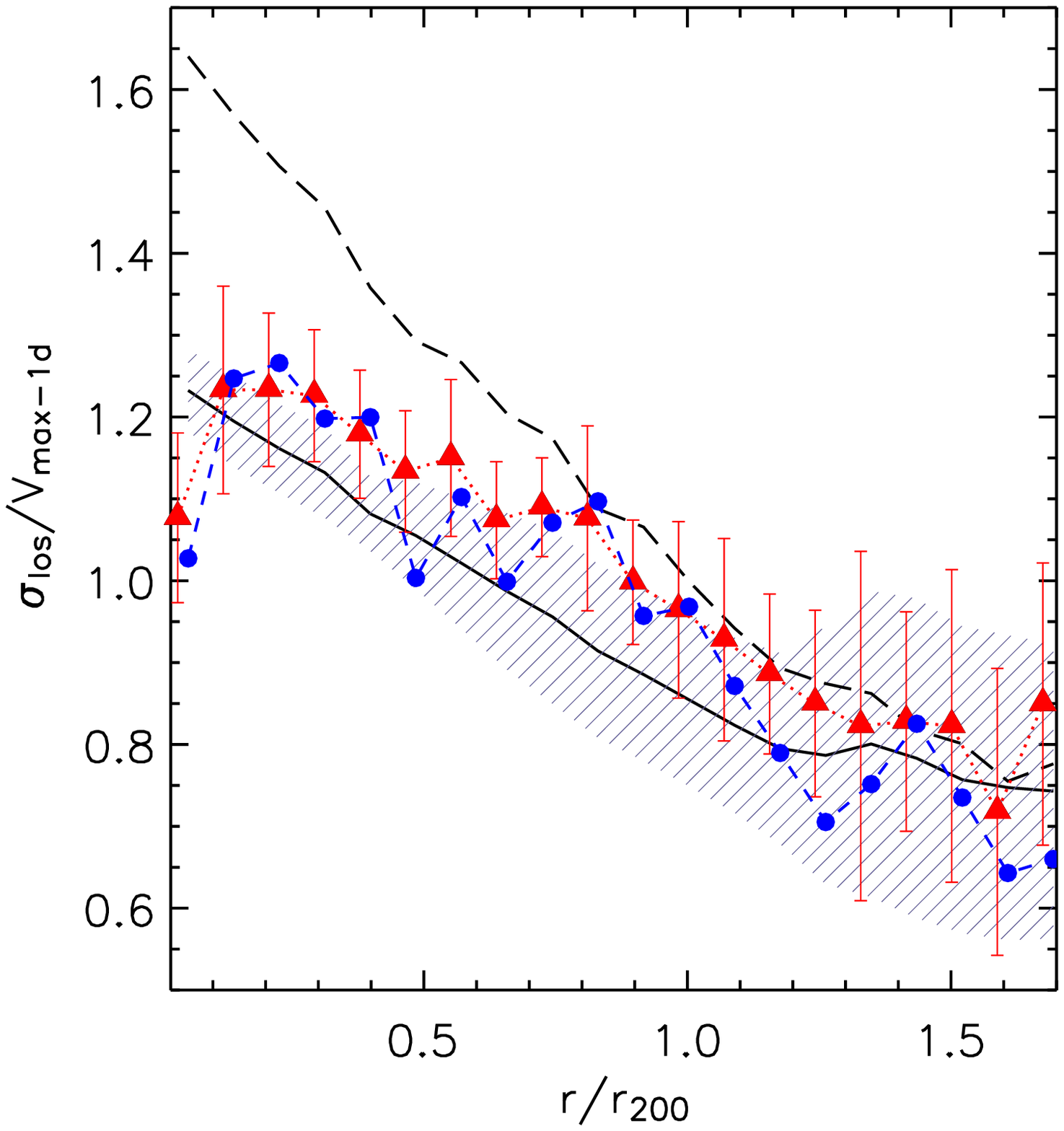}}
\caption{Top left: mean radial profiles for the dark matter (solid line), for
  model galaxies to two different magnitude limits (filled symbols), and for
  different subhalo samples, based on the ten clusters used in this study.  Top
  right: mean projected surface density profiles for the dark matter (solid
  line) and for model galaxies to two different magnitude limits (dashed and
  dotted lines).  The filled symbols represent the mean observed surface
  density profile of cluster galaxies in the CNOC survey (Carlberg et
  al.1997;).  In these two panels, the hashed region represents the full
  scatter in dark matter profiles.  Bottom panels: 3-D velocity dispersion
  profile (left) and line--of--sight velocity dispersion profile (right) for
  dark matter (solid line), for subhaloes containing at least 30
  particles( dash line), and for model galaxies to two different
  limiting magnitudes. The hashed regions and the error bars represent
  the standard $1\sigma$ scatter in the dark matter and the
  galaxy(${\rm B}<-17$) velocity dispersion profiles,
  respectively.}
  \label{fig:fig1}
\end{figure*}

A number of recent studies have focussed on the radial distribution of
subhaloes within dark matter haloes (Ghigna et al. 2001; Stoehr et al. 2003;
De Lucia et al. 2004a; Diemand et al. 2004; Gao et al. 2004; Gill et
al. 2004). These papers all agree that the subhalo profile is shallower than
that of the underlying dark matter, and indeed their subhalo profiles are all
very similar. Our own results are shown in the top left panel of
Fig.~\ref{fig:fig1} in the form of average radial profiles for the dark matter
and for different subhalo samples within our $10$ cluster resimulations. Note
that there are roughly $50$ subhaloes per cluster with $M_{\rm sub}/M_{\rm
halo}>2\times10^{-4}$ or with $V_{\rm sub}/V_{\rm halo}>0.09$. (The two
velocities here are the maximum circular velocities of the subhalo and of the
cluster respectively.) There are about 350 subhaloes per cluster with more
than 30 particles, which is the limit to which Gao et al. (2004) considered
the subhalo distributions to be insensitive to resolution effects.  The hashed
region shows the scatter of the dark matter density profiles in our simulation
set.  Note that all densities have been normalised to the mean density inside
the virial radius. The weak concentration of the subhalo distribution relative
to that of the dark matter is evident for all our samples, although, as noted
by Gao et al. (2004), the profile depends on how the subhalo population is
defined (limited in mass or in circular velocity). In this same panel we
also plot mean profiles for our model galaxies to two different magnitude
limits. In contrast to the subhalo profiles and in agreement with Diaferio et
al. (2001) and Springel et al. (2001), these coincide very nicely with the mean
dark matter density profile.

In the top right panel of Fig.~\ref{fig:fig1} we plot the average projected
dark matter distribution together with the surface density profile of model
galaxies to two different magnitude limits. For comparison, we also plot the
average observed surface density profile for cluster galaxies in the CNOC
survey (Carlberg et al. 1997).  The surface density profiles for the
simulations are obtained by projecting along the $x$, $y$ and $z$ axes in
turn, keeping only dark matter particles and galaxies within $\pm 2 R_{200}$
of cluster centre in depth, and binning up the projected density profiles out
to a projected distance of $2R_{200}$. The plotted curves are then an average
over three projections of each of ten simulations.  The mean galaxy surface
density profiles of our simulations agree extremely well both with the
observational data and with the mean dark matter profile.

In the bottom panels of Fig.~\ref{fig:fig1} we show the 3-D (left panel) and
the line--of--sight (right panel) velocity dispersion profiles of dark matter
particles, of galaxies and of subhaloes containing at least 30 particles. The
hashed region and error bars represent the standard $1\sigma$ scatter
of the dark matter and galaxy (${\rm B}<-17$) profiles among our ten
resimulations. In agreement with previous studies (Ghigna et
al. 1998; Colin et al. 2000; Diemand et al. 2004; Gill et al. 2004b),
we find that the velocity dispersion of the subhalo population
substantially exceeds that of the dark matter, particularly in the
inner regions. On the other hand, there is at most a weak upward bias
in the velocity dispersions of the model galaxies.

Fig.~\ref{fig:fig1} clearly shows that subhaloes and galaxies have very
different number density and velocity dispersion profiles in our simulations,
despite the fact that we assume that a galaxy forms at the centre of each dark
halo and is carried along with it when it falls into a larger system and so
becomes a subhalo. What is the origin of these differences?  If one wants to
relate the properties of subhaloes to those of the galaxies residing within
them, the evolution of the baryonic component has to be tracked
appropriately. This necessarily involves consideration of the full
collapse, assembly, merging and tidal stripping history of each
subhalo, rather than just its properties at the final time. Such
tracking can be carried out conveniently and moderately realistically
using semi--analytic techniques, as is done in this work.

Note that many of the model galaxies used to construct Fig.~\ref{fig:fig1} are
not associated with any resolved dark matter subhalo. In pure dark matter
simulations, subhaloes can disappear once their mass falls below the resolution
limit of the simulation. It may be that their dark matter content should
indeed be reduced to such small values by tidal stripping, or it may be that
proper inclusion of the effects of the baryonic component would make them more
resistant to stripping and disruption, as originally envisaged by White \&
Rees (1978). Our semi--analytic model assumes that the visible galaxy survives
even if the mass of the correponding subhalo drops below the limit of our
$N$--body simulation.  We associate the galaxy with the most bound particle of
its subhalo at the last time this could be identified, and we use this
particle at later times to track the galaxy's position and velocity. Such
`orphan' galaxies behave as individual $N$--body particles although we assume
them to merge with the central galaxy of the cluster on a dynamical friction
time--scale. At the resolution of our simulations, there are a
substantial number of these `orphan' galaxies and they are responsible
for the large differences between the `galaxy' and `subhalo' profiles
in the inner regions of our clusters. We demonstrate this by plotting the
number density profile for all galaxies brighter than ${\rm B}=-17$
which are still associated with subhaloes (i.e. not `orphan'
galaxies) as thin solid line in the top left panel of
Fig.~\ref{fig:fig1}. Clearly these galaxies are substantially less
concentrated than the galaxy population as a whole.

\begin{figure*}
\resizebox{8.5cm} {!}{\includegraphics{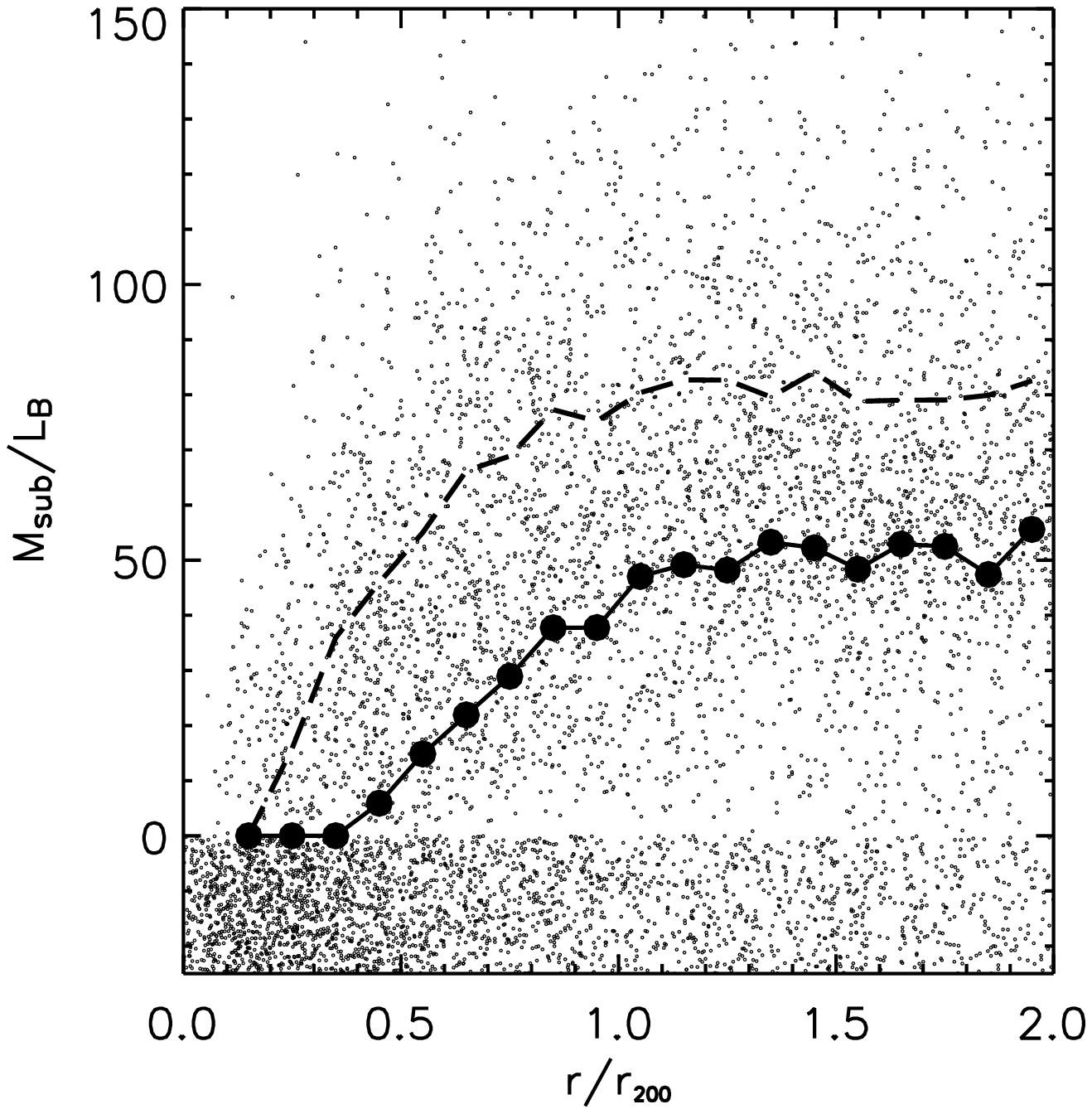}}\hspace{0.0cm}%
\resizebox{8.5cm}{!}{\includegraphics{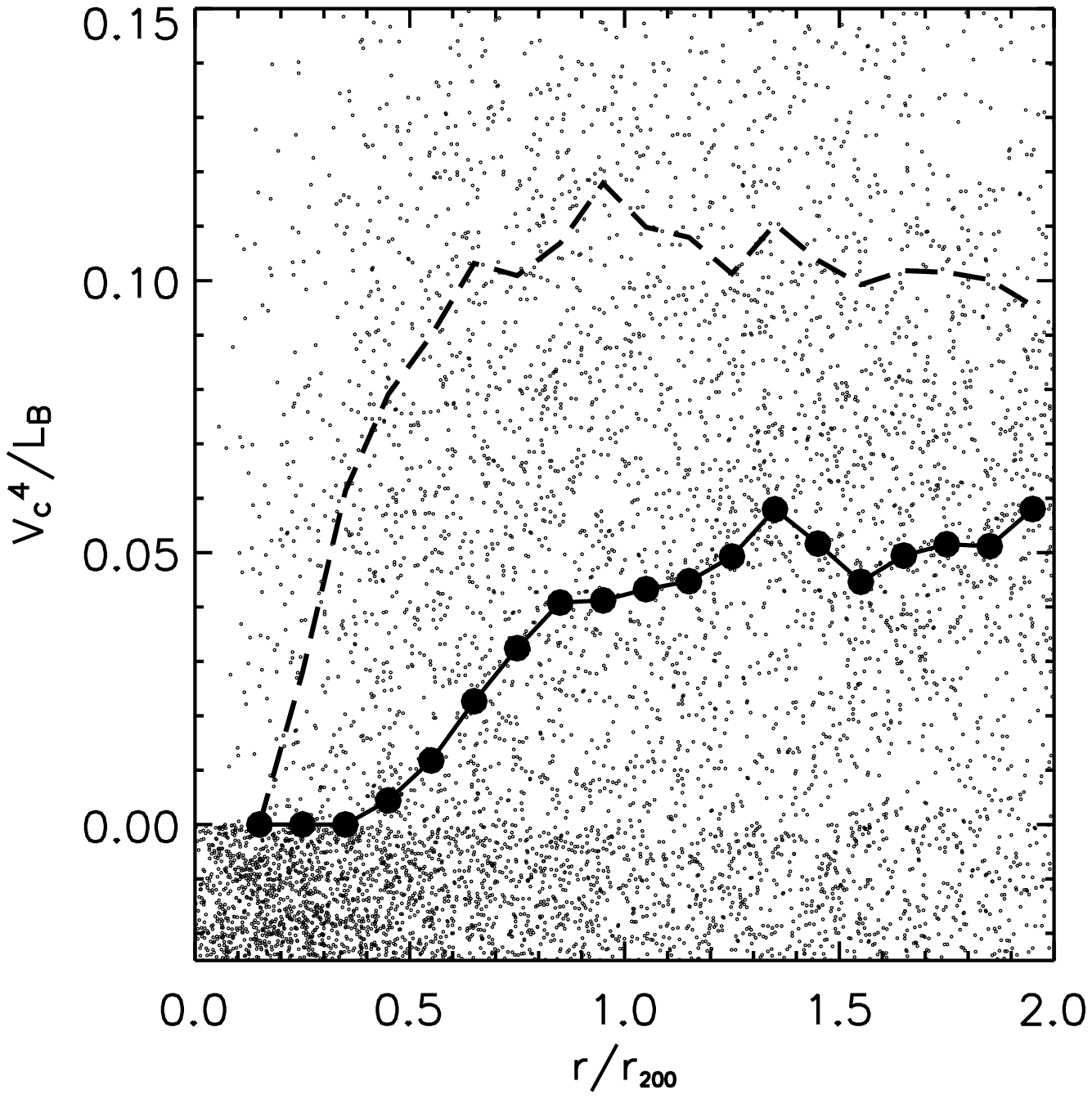}} \caption{Mass--to--light
ratio (left) and (circular velocity)$^4$--to--light ratio (right) for model
galaxies brighter than ${\rm B}=-17$ as a function of distance from
cluster centre. Filled cycles connected by thick lines show median
values as a function of radius; Dashed lines show the $80$th
percentile of the distribution. Galaxies that are not associated with
any dark matter subhalo are assigned zero mass and circular velocity,
but are displayed with a randomly generated small negative value of
the ordinate so that they are visible in the plots.}
\label{fig:fig2}
\end{figure*}

As discussed in Gao et al. (2004), the infall time and the retained mass of a
subhalo are both strongly increasing functions of clustercentric radius. This
implies that subhaloes in the inner regions of cluster haloes today were
generally more massive in the past than similar mass but more recently accreted
subhaloes in the outer regions.  As first shown by Springel et al. (2001), this
produces an increasing mass--to--light ratio as a function of the
clustercentric distance. We show this for our present models in
Fig.~\ref{fig:fig2} where we plot mass--to--light ratio ($M/L_{\rm B}$) and
(circular velocity)$^4$--to--light ratio ($V^{4}/L_{\rm B}$) for our
model galaxies as a function of distance from cluster centre. Galaxies
brighter than $M_{\rm B}=-17$ from all ten resimulations are shown
here.  The velocity used in the right panel is defined as the maximum
circular velocity of the associated subhalo. Outside the virial
radius, these ratios are almost flat, reflecting the proportionality
between the halo mass (or circular velocity to the fourth power) and
the galaxy luminosity for isolated haloes (the Tully--Fisher relation).  

About $55$ per cent of the cluster galaxies brighter than
$M_{\rm B}=-17$ are not associated to any resolved subhalo, and so are
assigned zero mass and circular velocity.  In order to show the
density distribution of these `orphan' galaxies in
Fig.~\ref{fig:fig2}, we assign them a mass--to--light ratio varying
randomly between $-25$ and $0$ and a (velocity)$^4$--to--light ratio
between $-0.025$ and $0$. Springel et al. (2001) show that this
$M/L_{\rm B}$ trend is present at a similar level in a simulation with
almost ten times better resolution than those we use here. In
addition, numerical convergence studies by Diemand et al. (2004) and
by Gao et al. (2004) indicate that resolution effects on subhalo mass
are relatively small for subhaloes with more than 30 particles and so
cannot be responsible for the trends in Figure 2. 

The radial variation of the mass--to--light ratio of cluster galaxies reflects
the fact that tidal stripping is very efficient in reducing the masses of
subhaloes within larger systems but is assumed to have much less effect on the
luminosity and structure of the galaxies which reside at their centres.  In
such a situation, selecting subhaloes above a certain mass (or circular
velocity) results in a population with very different properties from a galaxy
population selected above a certain limiting magnitude.
\section{Summary and discussion}
\label{sec:summary}
In this Letter, we have implemented a semi--analytic treatment of galaxy
formation on ten high resolution resimulations of galaxy cluster evolution in
order to study the number density and velocity dispersion profiles predicted
for galaxies and for dark matter subhaloes in $\Lambda$CDM galaxy clusters.
In agreement with previous work, we find galaxy profiles that agree well both
with simulated dark matter profiles and with observed galaxy profiles, but
subhalo profiles with much weaker central concentration and with substantially
higher velocity dispersion.

We show that these differences are due to a strong increase in the
mass--to--light (or (circular velocity)$^4$--to--light) ratio of galaxies as a
function of the distance from cluster centre. This trend is caused by tidal
stripping which rapidly reduces the mass of dark matter subhaloes once they
are accreted onto a larger structure, while only weakly affecting the galaxies
at their centres. In related work, De Lucia et al. (2004a) and Gao et
al. (2004) examine in considerably more detail the efficiency of tidal
stripping, showing that the longer a substructure spends in a massive halo,
the larger is the destructive effect.  As they demonstrate explicitly,
subhaloes are constantly being erased and being replaced by newly infalling
haloes.  Our semi--analytic model assumes that this process does not, however,
destroy the galaxy at the centre of each subhalo, which has typically
accumulated a substantial and strongly bound stellar component during earlier
evolutionary stages.

Much of the work on substructure within dark matter haloes has attempted to
link simulated substructure to observed galaxies by assuming a constant
mass--to--light ratio for subhaloes or by relating their maximum circular
velocity to galaxy luminosity through the observed Tully--Fisher and
Fundamental Plane relations. Our results show clearly, as did the earlier
results of Springel et al. (2001), that such assumptions are very unlikely to
give realistic results.  Galaxies and subhaloes are not simply related. The
luminosity of a galaxy cannot be inferred from the $z=0$ properties of the
subhalo which corresponds to it in a dark--matter--only $N$--body
simulation. Indeed, many cluster galaxies have {\it no} corresponding subhalo
in such a simulation, even though the haloes in which they originally formed
were easily resolved by the simulation. The galaxy formation process must be
treated appropriately to get results which are even qualitatively correct.

We note that these issues will not be addressed by carrying out dark
matter simulations of higher resolution. The tests of Diemand et
al. (2004) and Gao et al. (2004) show that subhaloes can be followed
and their masses tracked at least roughly down to a a limit of 20
particles or so, corresponding to subhalo masses around $10^{10}{\rm
M_\odot}$ for the simulations in this paper. This is {\it below} the
observed stellar mass of the galaxies in the real samples with which
we are comparing our models. Thus dynamical evolution becomes
dominated by the visible components of galaxies before our simulations
run into resolution problems. Any improvement over our current simple
semi--analytic assumptions will require explicit modelling of
structure in the stellar component of cluster galaxies. 

Finally we note that although this paper has dealt with cluster--sized 
haloes only, the same caveats apply also to galaxy-- and group--sized 
haloes. Only through a full treatment of the baryonic physics, is it possible 
to carry out a detailed comparison between theoretical results and
observational data.  A complex network of actions and back--reactions
regulates the evolution of the galaxy components we see, and any
comparison of simulated subhaloes to observed galaxies must consider the
time--integrated effect of these processes or risk serious error.

\section*{Acknowledgements}
The simulations used in this paper were carried out on the Cosma supercomputer
of the Institute for Computational Cosmology at the University of Durham. We
thank Volker Springel, Yipeng Jing and Xi Kang for useful
discussions. G.~D.~L. thanks the Alexander von Humboldt Foundation,
the Federal Ministry of Education and Research, and the Programme for
Investment in the Future (ZIP) of the German Government for financial
support.
\label{lastpage}

\end{document}